%% file: sadough_EW_HAL.tex
\documentclass[twocolumn]{IEEEtran}

\usepackage{amsmath,epsfig,amsfonts,amssymb,epsf,mathrsfs}
\usepackage{psfrag}

\title{Wavelet Based Semi-blind Channel Estimation For Multiband OFDM}

\author
   {Sajad Sadough$^{\dag *}$, Mahieddine M. Ichir$^{*}$, Emmanuel Jaffrot$^{\dag}$ and Pierre Duhamel$^{*}$\\ 
         $^{\dag}$ UEI, ENSTA, 32 boulevard Victor, 75015 Paris, France \\
         $^*$ Laboratoire des Signaux et Syst\`{e}mes, CNRS-Sup\'{e}lec,\\ Plateau de Moulon, 91190 Gif--sur--Yvette, France\\
         Email: \, sajad.sadough@ensta.fr}

\input abrmath.tex

\def\yb{\mathbf{y}}
\def\Yb{\mathbf{Y}}
\def\sb{\mathbf{s}}
\def\Sb{\mathbf{S}}

\def\hid{\widetilde{\Hb}}

\def\hb{\mathbf{h}}

\def\Hb{\mathbf{H}}
\def\gb{\mathbf{g}}
\def\gtb{\widetilde{\mathbf{g}}}
\def\gt{\widetilde{g}}
\def\Esp{\mathbb{E}_{\mathbf{S},\hid}}

\def\prob#1{p\left(#1\right)}
\def\ppi#1{\pi\left(#1\right)}
\def\Nc{\mathcal{N}}
\def\Ib{\mathbf{I}}
\def\Lt{\widetilde{L}}

\begin{document}
\maketitle

\begin{abstract}
%Multiband OFDM is a promising technology for high data-rate UWB communications.
This paper introduces an {\it expectation-maximization} (EM)
algorithm within a wavelet domain Bayesian framework for semi-blind
channel estimation of multiband OFDM based UWB communications. A
prior distribution is chosen for the wavelet coefficients of the
unknown channel impulse response in order to model a sparseness
property of the wavelet representation. This prior yields, in {\it
maximum a posteriori} estimation, a thresholding rule within the EM
algorithm. We particularly focus on reducing the number of estimated
parameters by iteratively discarding ``unsignificant'' wavelet
coefficients from the estimation process. Simulation results using
UWB channels issued from both models and measurements show that under sparsity conditions,
the proposed algorithm outperforms pilot based channel estimation in terms of mean square error and bit
error rate and enhances the estimation accuracy with less
computational complexity than traditional semi-blind methods.
\end{abstract}
%##############################################################################
%        Section 1:   Introduction
%##############################################################################
\section{Introduction}
\label{sec:intro}
%\No PARstart
%Ultra-Wide-Band (UWB) is a wireless technology for high data rate, short range transmission over the 3.1-10.6 GHz frequency band.
A UWB radio signal is defined as any signal whose bandwidth is larger than 20\% of its center frequency or greater than 500 MHz \cite{fcc}.
%UWB systems deal with signals with a fractional bandwidth that is larger than 25 \% of the center frequency or greater than 500 MHz \cite{fcc}.
In recent years, UWB system design has experienced a shift from the traditional ``single-band''
radio that occupies the whole 7.5 GHz allocated spectrum to a ``multiband'' design approach \cite{Roy2004a}.
That consists in dividing the available UWB spectrum into
several subbands, each one occupying approximately 500 MHz.

Multiband Orthogonal Frequency Division Multiplexing (MB-OFDM)
\cite{batra_jour} is a strong candidate for multiband UWB which enables
high data rate UWB transmission to inherit all the strength of OFDM that has
already been shown for wireless communications (ADSL, DVB, 802.11a,
802.16.a, etc.). This approach uses a conventional coded OFDM system \cite{prasad1} together with bit interleaved coded modulation
(BICM) and frequency hopping over different subbands to improve diversity and to enable multiple access.

Basic receivers proposed for MB-OFDM \cite{batra_jour}, estimate the
channel by using pilots (known training symbols) transmitted at the
beginning of the information frame, implicitly assuming a time
invariant channel within a single frame. Thus, for an accurate channel acquisition, one must send several pilot patterns resulting in a significant loss in spectral efficiency.

Recent works \cite{Kobayashi2004a, Naffouri2002b} have reported promising results on the combination of
channel estimation and data decoding process by using the Expectation-Maximization (EM) algorithm \cite{EM2} . 
Though the latter scheme outperforms
pilot based receivers, it has a higher complexity
that may be of a critical concern for its practical implementations. This complexity is
mainly dominated by the number of estimated parameters for channel updating and the decoding algorithm within each iteration.

In this work, we consider a semi-blind joint channel estimation and
data detection scheme based on the EM algorithm, with the objective
of minimizing the number of estimated parameters and enhancing the
estimation accuracy. This is achieved by expressing the unknown
channel impulse response (CIR) in terms of its discrete wavelet
series, which has been shown to provide a {\it parsimonious}
representation \cite{silver1,sad1}. Thus, we choose a particular
prior distribution for the channel wavelet coefficients that renders
the maximum a posteriori (MAP) channel estimation equivalent to a
hard thresholding rule at each iteration of the EM algorithm. The
latter is then exploited to reduce the estimator computational load
by discarding ``unsignificant'' wavelet coefficients from the
estimation process.
Moreover, since the probability of encoded bits are involved in the EM computation,
we naturally combine the iterative process of channel estimation with the decoding operation of encoded data.

This paper is organized as follows.
Section \ref{obs_model} introduces MB-OFDM and its wavelet domain channel estimation observation model.
In section \ref{emMap}, we first describe a MAP version of the EM algorithm for channel estimation and then show how the number of estimated parameters can be reduced through the EM iterations. The combination of the channel estimation part with the decoding operation and implementation issues are also discussed. Section \ref{simul} illustrates, via simulations, the performance of the proposed receiver over a realistic UWB channel environment and section \ref{concl} concludes the paper.

Notational conventions are as follows: $\mathcal{D}_{\mathbf{x}}$ is a diagonal matrix
with diagonal elements $\mathbf{x}=[x_1,\ldots,x_N]^T$, $\mathbb{E}_{\mathbf{x}}[.]$ refers to expectation
with respect to $\mathbf{x}$, $\mathbf{I}_N$ denotes an $(N \times N)$ identity matrix; $\|.\|$, $(.)^*$, $(.)^T$
and $(.)^{\mathcal{H}}$ denote Frobenious norm, matrix or vector conjugation, transpose and Hermitian transpose, respectively.
%###########################################################################
%                              section MB-OFDM overview
%###########################################################################
\section{System model and wavelet domain problem formulation}
\label{obs_model}

\begin{figure}[!htb]
\centering
%\includegraphics[width=0.45\textwidth,height=4cm]{TFC_e}
%\psfrag{A}{
\begin{tabular}{@{}ll}
  \begin{tabular}{@{}c@{}}
    \rotatebox{90}{\hspace{-3.5em}{\bf 3 lower subbands}}
    $\left\{ \begin{array}{c}
        \\    \\
        \\    \\
        \\    \\
        \\    \\
      \end{array} \right.$
  \end{tabular}
  &
  \begin{tabular}{@{}c@{}}
    \psfrag{B}{\hspace{-2.2em}\scriptsize \raisebox{-1em}{Band \# 1}}
    \psfrag{C}{\hspace{-2.2em}\scriptsize \raisebox{-1em}{Band \# 2}}
    \psfrag{D}{\hspace{-2.2em}\scriptsize Band \# 3}
    \psfrag{1}{{\scriptsize\hspace{-1em}3168}}
    \psfrag{2}{\scriptsize\hspace{-1em}3696}
    \psfrag{3}{\scriptsize\hspace{-1em}4224}
    \psfrag{4}{\scriptsize\hspace{-1em}4752}
    \psfrag{E}{\hspace{-1em}\scriptsize  9.5 ns}
    \psfrag{F}{\hspace{-1.5em} \parbox{8.5em}{\centering\scriptsize Guard Interval for\\ TX/RX switching}}
    \psfrag{G}{\scriptsize 312.5 ns}
    \psfrag{I}{\hspace{-.5em}\scriptsize 60.6 ns}
    \psfrag{A}{\hspace{-2em}\scriptsize Time (ns)}
    \psfrag{H}{\hspace{-.5em}\rotatebox{90}{\hspace{-4em} Frequency (MHz)}}
    \includegraphics[width=0.4\textwidth,height=4cm]{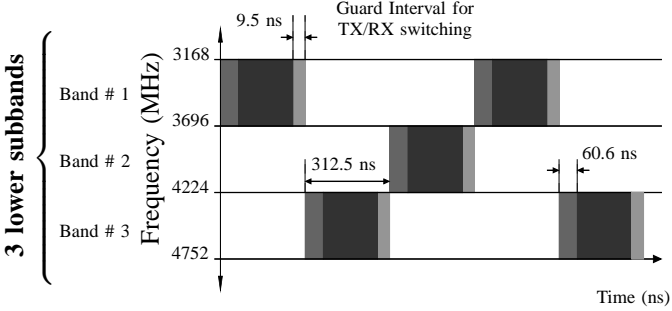}
  \end{tabular}
\end{tabular}
\caption{Example of time-frequency coding for the multiband OFDM system: TFC=\{1, 3, 2, 1, 3, 2, ...\}.} \label{TFC}
\end{figure}

\begin{figure}[!t]
\centering
\psfrag{A}{\hspace{-1.5em}{\small Binary Data}}
\psfrag{B}{\hspace{-.5em}{\parbox{5.5em}{\centering {\small Punctured Convolutional Encoder}}}}
\psfrag{C}{\hspace{-1.2em}{\parbox{6em}{\centering {\small Bit\\ Interleaver}}}}
\psfrag{D}{\hspace{-1.2em}\parbox{6em}{\centering {\small QPSK\\ Mapping}}}
\psfrag{E}{\hspace{-1.2em}\parbox{6em}{\centering {\small IFFT\\ CP \& GI\\ Addition}}}
\psfrag{F}{\hspace{-2em}\parbox{6em}{\centering {\small DAC}}}
\psfrag{G}{\hspace{-3em}\parbox{6em}{\centering $\exp\left(j2\pi f_ct\right)$}}
\psfrag{H}{\hspace{-.5em}{{\small TFC: Subband Selection}}}
\includegraphics[width=0.45\textwidth,height=4cm]{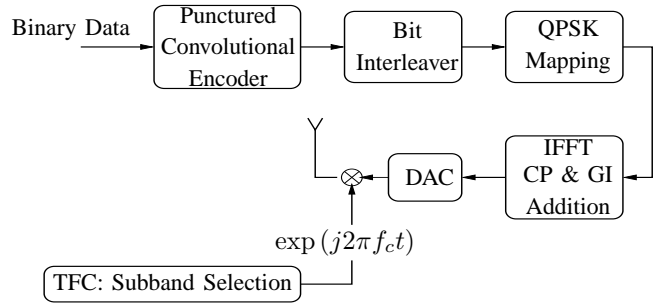}
\caption{TX architecture of the multiband OFDM system.}\label{fig1}
\end{figure}

MB-OFDM system divides the spectrum between
3.1 to 10.6 GHz into several non-overlapping subbands each one
occupying 528 MHz of bandwidth \cite{batra_jour}.
The transmitter architecture for
the MB-OFDM system is very similar to that of a conventional
wireless OFDM system. The main difference is that MB-OFDM system
uses a time-frequency code (TFC) to select the center frequency of
different subbands which is used not only to provide frequency
diversity but also to distinguish between multiple users (see figures
\ref{TFC} and \ref{fig1}). 
Here, we consider MB-OFDM
in its basic mode {\it ie.} employing the three first subbands. 

We consider the multiband OFDM transmission of figure \ref{fig1} using $N$ data subcarriers.
At the receiver, assuming a cyclic prefix (CP) longer than the channel maximum delay spread and perfect synchronization,
OFDM converts a frequency selective channel into $N$ parallel flat fading subchannels \cite{prasad1} for each subband as
\begin{equation}
\yb_{i,n} = \mathcal{D}_{{\sb_{i,n}}} \, \hb_{i,n} + \mathbf{z}_{i,n}  \;\;\;\;i\in\{1,2,3\},\;\;\; n=1,\ldots,{\rm N_{sym}}
\end{equation}
where $(1\times N)$ vectors $\yb_{i,n}$, $\sb_{i,n}$ and $\hb_{i,n}$ denote received and transmitted symbols,
and the channel frequency response respectively;  the noise block $\mathbf{z}_{i,n}$ is assumed
to be a zero mean white complex Gaussian noise with distribution $\mathcal{CN}(\mathbf{0},\sigma^2\mathbf{I}_N)$ ; $i$ is
the subband index and $n$ refers to the OFDM symbol index inside the frame. The observation model corresponding to all
three subbands can be written in frequency domain as
\begin{equation}
\label{obs3B}
  \Yb_m = \mathcal{D}_{_{\Sb_m}} \, \Hb_m + \mathbf{Z}_m \;\;\;\;m=1,\ldots,{\rm M_{sym}}
\end{equation}
where $\Yb_m={[\yb_{1,n}, \yb_{2,n}, \yb_{3,n}]^T}$, $\Sb_m={[\mathbf{s}_{1,n}, \mathbf{s}_{2,n}, \mathbf{s}_{3,n}]^T}$, $\Hb_m={[\hb_{1,n}, \hb_{2,n}, \hb_{3,n}]^T}$ and $\mathbf{Z}_m={[\mathbf{z}_{1,n}, \mathbf{z}_{2,n}, \mathbf{z}_{3,n}]^T}$ are ($M \times 1$) vectors, with $M=3N$ and ${\rm M_{sym}}={\rm N_{sym}}/3$. In the remainder, unless otherwise mentioned, we will not write the time index $m$ for notational convenience.

In order to take advantage of the wavelet based estimation, the channel impulse
response is expressed in terms of its orthogonal discrete wavelet coefficients. Let $\mathbf{F}_{M,L}$ be the truncated
fast Fourier transform (FFT) matrix constructed from the ($M\times M$) FFT matrix by keeping the first $L$ columns where $L$
is the length of the CIR over a group of three subbands. We define $\mathbf{W}$ as the $(L\times L)$ orthogonal discrete wavelet transform (ODWT) matrix.
The unknown channel can be expressed as $\Hb=\mathbf{F}_{M,L} \mathbf{W}^{\mathcal{H}} \mathbf{g}$, where $\mathbf{g}$ is the $(L \times 1)$
vector of the CIR wavelet coefficients. The Observation model \ref{obs3B} is rewritten as
\begin{equation}
   \label{eq:obs2}
  \Yb=\mathcal{D}_{_\Sb}\,\mathbf{T}\,\mathbf{g} + \mathbf{Z}
\end{equation}
where $\mathbf{T}=\mathbf{F}_{M,L}\mathbf{W}^{\mathcal{H}}$.

Although at the transmitter, the channel is practically used by
slices of 528 MHz bandwidth that corresponds to one of the subbands,
at the receiver side we gather three received OFDM symbols for
estimating the wavelet coefficients of the CIR, taken over all of
the subbands (1.584 GHz bandwidth). This is motivated by the fact
that estimating the channel over a wider bandwidth leads to a
sparser wavelet representation. Besides, this approach
simplifies the receiver architecture since there is no need to
change the central frequency for down converting different subbands.
%#############################################################################
%                               section EM-MAP Algorithm
%#############################################################################
\section{The EM-MAP algorithm for wavelet domain channel estimation}
\label{emMap}
The EM algorithm proposed in this section is able to integrate the advantages of wavelet based estimation via the prior choosen for channel wavelet coefficients. Next, we see how the MAP estimator leads to a thresholding procedure which is used for reducing the number of estimated coefficients at each iteration of the EM algorithm.
%*******************************subsection 4.1***********************************
\subsection{An equivalent model and the EM principle}
%********************************************************************************
Our first step consists in decomposing the AWGN in \eqref{eq:obs2} into the sum of two different Gaussian noise terms as
\begin{equation}
  \mathbf{Z} = \mathcal{D}_{_\Sb}\mathbf{Z}_1 + \mathbf{Z}_2
\end{equation}
where $\mathbf{Z}_1$ and $\mathbf{Z}_2$ are $(M\times1)$ independent
Gaussian noise vectors such that
$p(\mathbf{Z}_1)=\mathcal{CN}(\mathbf{0},\alpha^2 \mathbf{I}_M)$ and
$p(\mathbf{Z}_2)=\mathcal{CN}(\mathbf{0},\sigma^2
\mathbf{I}_M-\alpha^2 \mathcal{D}_{_\Sb}
\mathcal{D}^{\mathcal{H}}_{_\Sb})$. Since we are using normalized
QPSK symbols,
$\mathcal{D}_{_\Sb}\mathcal{D}^{\mathcal{H}}_{_\Sb}=\mathbf{I}_M$
and the covariance matrix of $\mathbf{Z}_{\mathrm{2}}$ reduces to
$\mathbf{\Sigma}_{_2} = (\sigma^2-\alpha^2) \mathbf{I}_M$. We define
the positive design parameter $\rho\triangleq\alpha^2/\sigma^2$, $(0 <
\rho\leq 1)$ and notice that setting $\rho =1$ leads to
$\mathbf{Z}_2=0$ which is equivalent to working with the initial
model \eqref{eq:obs2}. However, for $0<\rho< 1$, the above noise
decomposition allows the introduction of a hidden channel vector
$\hid$ defined as
\begin{equation}
   \label{obs2step}
     \left\{\begin{array}{ll}
      \hid &= \mathbf{T} \, \mathbf{g} + \mathbf{Z}_1 \\
      \Yb &= \mathcal{D}_{_\Sb} \, \hid + \mathbf{Z}_2.
    \end{array} \right.
\end{equation}
The hidden vector $\hid$ provides a direct relation between true and
estimated wavelet coefficients corrupted by an additive white
Gaussian noise, allowing the two-stage observation model
\eqref{obs2step} which is equivalent to \eqref{eq:obs2}. However,
the difference with a standard denoising problem 
is that $\Sb$ and $\hid$ are unknown. Hence, the
observation model has missing datas and hidden variables and the MAP
solution of $\mathbf{g}$ has no closed form. In such situations, the
EM algorithm \cite{EM2} is often used to maximize the expectation
of the posterior distribution over all possible missing and hidden
variables.

Let $\mathbf{X}=\{\Yb,\Sb,\hid \}$ be the {\it complete data set} in
the EM algorithm terminology. Note that the observation set $\Yb$
determines only a subset of the space $\mathscr{X}$ of which
$\mathbf{X}$ is an outcome. We search $\mathbf{g}$ that maximizes
$\log p(\mathbf{g}|\mathbf{X})$. After initialization by a short pilot sequence at the beginning of the frame,
the EM algorithm alternates between the following two steps (until
some stopping criterion) to produce a sequence of estimates $\{
\mathbf{g}^{(t)},\, t=0,1,\ldots,t_{\rm max} \}$.
\begin{itemize}
\item \textbf{Expectation Step} (E-step): The conditional expectation of
the complete log-likelihood given the observed vector and the current estimate $\mathbf{g}^{(t)}$ is
calculated. This quantity is called the {\it auxiliary} or $Q$-function
  \begin{equation}
    \label{Estep}
    Q\big(\mathbf{g},\mathbf{g}^{(t)}\big)=\Esp \Big[\log p(\mathbf{Y},\mathbf{S},\hid|\mathbf{g})\Big|\mathbf{y},\mathbf{g}^{(t)}\Big]
  \end{equation}

\item \textbf{Maximization Step} (M-step): The estimated parameter is updated according to
  \begin{equation}
  \label{Mstep}
 {\mathbf{g}}^{(t+1)}=\Argmax{\gb}{Q\big(\gb,{\gb}^{(t)}\big)+ \log \pi(\gb)}
  \end{equation}
\end{itemize}
where $\pi(\gb)$ is a {\it prior} distribution for the wavelet coefficients.
Next, we derive the specific formulas of each step, according to \eqref{obs2step}.
%******************** subsection 4.2 *******************
\subsection{E-step: Computation of the $Q$-function}
%*******************************************************
The complete likelihood is
\begin{equation*}
  p(\mathbf{Y},\mathbf{S},\hid|\gb)=p(\mathbf{Y}|\mathbf{S},\hid,\gb)\, p(\mathbf{S}|\hid,\gb)\, p(\hid|\gb).
\end{equation*}
According to \eqref{obs2step}, conditioned on $\hid$, $\mathbf{Y}$ is independent of $\gb$. Furthermore, $\mathbf{S}$
which results from coding and interleaving of bit sequence is independent of $\hid$ and $\gb$. Since $\mathbf{Z}_1$ is
a complex white Gaussian noise, the complete log-likelihood can be simplified to
\begin{eqnarray}
  \log p(\mathbf{Y},\mathbf{S},\hid|\gb)
  &=&\log \big[p(\mathbf{Y}|\mathbf{S},\hid)\;p(\mathbf{S})\;p(\hid|\gb)\big] \notag \\
%%%
  &=&\log p(\hid|\gb) \;+\; \textrm{cst.} \notag \\
%%%
  %&=&-\frac{{\|\, \hid-\mathbf{T}\gb\, \|}^2}{{\alpha}^2} \;+\; \textrm{cst.} \notag \\
                                %&=-\frac{{\mathbf{z}}^H \mathbf{z}-2{\boldsymbol{\theta}}^H \mathbf{T}^H \mathbf{z}+{\boldsymbol{\theta}}^H {\mathbf{T}}^H \mathbf{T} \boldsymbol{\theta}}{{\alpha}^2}+k^{'} \notag \\
  &=&-\frac{\gb^{\mathcal{H}} \mathbf{T}^{\mathcal{H}} \mathbf{T} \gb-2\gb^{\mathcal{H}} \mathbf{T}^{\mathcal{H}}\hid}{{\alpha}^2} \;+\; \textrm{cst.} \notag \\
\end{eqnarray}
where cst. are different constant terms that do not depend on $\gb$. According to \eqref{Estep} we have
\begin{align}
   \label{Qfunc}
  Q\big(\gb,{\gb}^{(t)}\big)&=\Esp \Big[ -\frac{\gb^{\mathcal{H}} \mathbf{T}^{\mathcal{H}} \mathbf{T} \gb-2\gb^{\mathcal{H}} \mathbf{T}^{\mathcal{H}} \hid}{{\alpha}^2} + \textrm{cst.} \, \Big|\mathbf{Y},\gb^{(t)} \Big] \notag \\
%%%
  %&=-\frac{\gb^{\mathcal{H}} \mathbf{T}^{\mathcal{H}} \mathbf{T} \gb-2\gb^{\mathcal{H}} \mathbf{T}^{\mathcal{H}} \, \Esp [\hid|\mathbf{Y},\gb^{(t)}]}{{\alpha}^2}  + \textrm{cst.}  \notag \\
%%%
  &=-\frac{{\|\, \langle \hid^{(t)} \rangle- \mathbf{T}\gb \, \|}^2}{{\alpha}^2} + \textrm{cst.}
\end{align}
where $\langle \hid^{(t)} \rangle \triangleq \Esp[\hid|\mathbf{Y},\gb^{(t)}]$.

From \eqref{Qfunc}, it is obvious that the E-step involves only the computation of $\langle \hid^{(t)} \rangle$, we have
\begin{align}
    \label{brace}
  %\langle \hid^{(t)} \rangle&=\Esp[\hid|\mathbf{y},\gb^{(t)}]    \notag \\
  %&=\sum_{\mathbf{S}\in \mathscr{C}} \int_{\hid \in \mathscr{H}} \hid \quad p(\hid,\mathbf{S}|\mathbf{Y},\gb^{(t)}) \, \mathrm{d} \hid \notag \\
%%%
  \langle \hid^{(t)} \rangle&=\sum_{\mathbf{S}\in \mathscr{C}} {\bigg(\int_{\hid \in \mathscr{H}} \hid \quad p(\hid|\mathbf{Y},\gb^{(t)}) \, \mathrm{d} \hid \bigg)} p(\mathbf{S}|\mathbf{Y},\gb^{(t)})
\end{align}
where the last equation results from the independence
between $\Sb$ and $\hid$ belonging respectively to the sets $\mathscr{C}$ and $\mathscr{H}$ which
contain all of their possible values. %Note that each entry of $\mathbf{S}$ takes one (unknown) discrete value inside the QPSK constellation whereas components of $\hid$ are continuous variables.

In order to evaluate $\langle \hid^{(t)} \rangle$, we first have to evaluate the conditional mean $\boldsymbol{\mu}_{\hid}^{(t)}$ of $\hid$ as
\begin{equation}
\boldsymbol{\mu}_{\hid}^{(t)}= \int_{\hid \in \mathscr{H}} \hid \quad p(\hid|\mathbf{Y},\gb^{(t)}) \, \mathrm{d} \hid
\end{equation}
Since both $p(\mathbf{Y}|\hid)$ and $p(\hid|\gb^{(t)})$ are Gaussian densities, $p(\hid|\mathbf{Y},\gb^{(t)})\propto p(\mathbf{Y}|\hid)\; p(\hid|\gb^{(t)})$
is also Gaussian. By standard manipulation of Gaussian densities, we obtain
\begin{equation}
\label{mean_g}
\boldsymbol{\mu}_h^{(t)}=\mathbf{T}\gb^{(t)}+ \rho\, \mathcal{D}_{_\Sb}^{\mathcal{H}}\Big(\mathbf{Y}-\mathcal{D}_{_\Sb} \mathbf{T}\gb^{(t)}\Big).
\end{equation}
By using \eqref{mean_g} in \eqref{brace} and after some simplifications we get
\begin{equation}
\label{estep_end}
\langle \hid^{(t)}\rangle=(1-\rho)\,\mathbf{T}{\gb}^{(t)}+\rho \,\overline{\mathcal{D}}^{\mathcal{H}}_{_\Sb}\mathbf{Y}
\end{equation}
where $\overline{\mathcal{D}}_{_\Sb}=\sum_{\mathbf{s}\in \mathscr{C}}\mathcal{D}_{_\Sb}\, p(\Sb|\Yb,{\gb}^{(t)})$.

The E-step is then completed by inserting $\langle \hid^{(t)} \rangle$ into $Q(\gb,\gb^{(t)})$, equation \eqref{Qfunc}.
%************************** subsection 4.3  ***********************
\subsection{M-step: Wavelet Based MAP Estimation}
%******************************************************************
In this step the estimate of the parameter $\gb$ is updated as given in \eqref{Mstep} where $Q(\boldsymbol{\theta},{\boldsymbol{\theta}}^{(t)})$ is given by \eqref{Qfunc}
\begin{equation}
\label{mstep1}
{\gb}^{(t+1)}=\Argmax{\gb}{-\frac{{\|\,\langle \hid^{(t)} \rangle-\mathbf{T}\gb\,\|}^2}{{\alpha}^2}+\log \pi(\gb)}.
\end{equation}
Due to the orthonormality of both Fourier and wavelet transforms, $\mathbf{T}^{\mathcal{H}} \mathbf{T}=\mathbf{I}_L$ and we can replace ${\|\,\langle\hid^{(t)} \rangle-\mathbf{T}\gb\,\|}^2$ by ${\|\,\gtb^{(t)}- \gb\,\|}^2$, where
\begin{align}
\label{mstep1_bis}
\gtb^{(t)}&= {\mathbf{T}}^{\mathcal{H}} \langle \hid^{(t)} \rangle \notag \\
          &= (1-\rho)\,{\gb}^{(t)}+\rho \,( \overline{\mathcal{D}}_{_\Sb} \mathbf{T})^{\mathcal{H}} \mathbf{Y}
\end{align}
The M-step can be written as
\begin{equation}
\label{mstep2}
\gb^{(t+1)}=\Argmax{\gb}{ -\frac{ {\|\, \gtb^{(t)}-\gb \,\|}^2 }{\alpha^2}+ \log \pi(\gb)}.
\end{equation}

Actually $\gb^{(t+1)}$ in \eqref{mstep1} is no more than the MAP estimate of $\gb$ from the observation model
\begin{equation}
\label{mstep3}
\gtb^{(t)} = \gb + \mathbf{Z}'_1
\end{equation}
where $\mathbf{Z}'_1 =\mathbf{T}^{\mathcal{H}} \mathbf{Z}_1 \sim\mathcal{CN}(\mathbf{0},\alpha^2\mathbf{I}_L)$.
From the Bayes theorem, the posterior distribution of $\gb$ is given by
\begin{equation}
  \prob{\gb|\gtb^{(t)}} \;\propto\; \prob{\gtb^{(t)}|\gb} \ppi{\gb}
\end{equation}
where $p(\gtb^{(t)}|\gb)$ is the Gaussian likelihood, $\gtb \sim \mathcal{CN}(\gb, \alpha^2\Ib_L)$. In this approach, we adopt the Bernoulli-Gaussian prior distribution $\ppi{\gb}$ for the wavelet coefficients $\gb$ of the unknown CIR described by
\begin{equation}
  \label{eq:BGmodel}
  \ppi{g_j} = \lambda \; \delta(g_j) + (1-\lambda)\; \mathcal{CN}_{g_j}\left(0,\tau^2\right)
\end{equation}
for $j = 1,\ldots,L$, which allows us to model a sparseness property of UWB channels in wavelet domain. This amounts
considering that the wavelet coefficients have a probability
$\lambda$ to be zero and a probability $1-\lambda$ to be
distributed as $\mathcal{CN}(0,\tau^2)$.
In order to deal with that particular model, we introduce an additional state variable (or indicator) $\beta_j \in \{0,1\}$
such that we can express this prior conditionally as
\begin{equation}
  \left\{
  \begin{array}{lllll}
    (g_j|\beta_j = 0) &\sim& \delta(g_j) & \textrm{with probability}\; \lambda, \\ \\
    (g_j|\beta_j = 1) &\sim& \mathcal{CN}_{g_j}\left(0,\tau^2\right) &\textrm{with probability}\; 1-\lambda. \\
  \end{array}
  \right.
\end{equation}
This prior model, conditionally on that state variable, leads to a
Gaussian posterior for $g_j$ which makes the estimation explicit;
from the direct observation model $\gt_j^{(t)} = g_j +  Z'_{1,j}$,
we can express these posterior probabilities of $\beta_j$ as
\begin{equation}
  \begin{array}{lll}
    \prob{\beta_j = 0|\gt_j^{(t)}} &=& \lambda \; \mathcal{N}\left(0, \alpha^2\right) /c \\
    \prob{\beta_j = 1|\gt_j^{(t)}} &=& (1-\lambda) \; \Nc\left(0, \alpha^2 + \tau^2\right) /c
  \end{array}
\end{equation}
where the constant $c = \lambda \, \mathcal{N}\left(0, \alpha^2\right) + (1-\lambda) \, \mathcal{N}\left(0, \alpha^2 + \tau^2\right)$.
From this set of equations, we easily notice that the indicator variable $\beta_j$ allows us to discriminate between the
noise coefficients (for $\beta_j=0$) and the effective channel wavelet coefficients (for $\beta_j=1$),
eventually corrupted by noise. The indicator variables $\beta_j$ are estimated, in the MAP sense, by
\begin{equation}
   \label{mstep4}
  \beta_j^{(t+1)} = \left\{
  \begin{array}{ll}
    0, & \textrm{if } \; \prob{\beta_j=0|\gt_j^{(t)}} \geq 0.5 \\ \\
    1, & \textrm{elsewhere}.
  \end{array}
  \right.
\end{equation}
Therefore, the MAP estimates of the channel wavelet coefficients are obtained by a simple denoising/thresholding rule as
\begin{equation}
   \label{mapstep5}
  g_j^{(t+1)} = \left\{
  \begin{array}{ll}
    0, &\textrm{if }\; \beta_j^{(t+1)} = 0 \\
    \displaystyle \frac{\tau^2}{\alpha^2 + \tau^2} \,\, \gt_j^{(t+1)}, &\textrm{if }\; \beta_j^{(t+1)} = 1 %& \textrm{elsewhere}.
  \end{array}
  \right.
\end{equation}
\\
%%%%%%%%%%%%%%%%%%%%%%%%%%%%%%%%%%%%%%%%%%%%
\subsubsection{$\tau$ and $\lambda$ updating}
%%%%%%%%%%%%%%%%%%%%%%%%%%%%%%%%%%%%%%%%%%%%
The prior parameters $\tau$ and $\lambda$ stand respectively for the (significant)-wavelet coefficients energy and unsignificant coefficient probability. The update rules for these two parameters are MAP based rules derived from assigning conjugate priors to these parameters \cite{Bernardo1994a}:
\begin{equation}
\label{parUpdate}
   \begin{array}{lll}
      \hat{\lambda} &=& {(\Lt - 1/2)}/{(L - 1)}, \\
      \hat{\tau}^2  &=& {\eta}/{(L - \Lt)} 
   \end{array} 
\end{equation}
where $\Lt = \textrm{Card}\{j \; \big| \;\beta_j = 0\}$ and $\eta = \sum_{\beta_j = 1} \big|g_j^{(t+1)}\big|^2$; \textrm{Card}\{.\} denoting the set cardinality.
\\
%%%%%%%%%%%%%%%%%%%%%%%%%%%%%%%%%%%%%%%%%%%%%%%%%%%%%%%%%%%%%%%
\subsubsection{Reduction of the number of estimated parameters}
%%%%%%%%%%%%%%%%%%%%%%%%%%%%%%%%%%%%%%%%%%%%%%%%%%%%%%%%%%%%%%%
\label{noise_reduc}
The thresholding procedure derived in this section, provides an easy
framework for reducing the number of estimated coefficients. This
can be done by discarding at each iteration, the elements of
${\gb}^{(t+1)}$ that are replaced by zero in \eqref{mapstep5}. The
underlying assumption is as follows: whenever the estimator
attributes an unknown wavelet coefficient to noise (replace it by
zero), this coefficient will always be considered as noise and so
will not be estimated in future iterations. 

This operation is shown in figure \ref{fig2} and can be modeled as:
\begin{equation}
 \label{mstep5}
\gb^{(t+1)}_{\mathrm{tr}} = \Theta \big(\, {\gb}^{(t+1)}\, \big), \quad \mathbf{T}_{\mathrm{tr}}= \Xi\big(\, \mathbf{T}\, \big)
\end{equation}
where the truncation operator $\Theta(.)$ gathers in $\gb^{(t+1)}_{\mathrm{tr}}$
the components of $\gb^{(t+1)}$ that must be kept and the operator $\Xi(.)$ constructs
$\mathbf{T}_{\mathrm{tr}}$ from $\mathbf{T}$ by keeping the rows corresponding to kept indexes.
During the first iteration ($t=0$), the algorithm does not perform any truncation and the EM algorithm
estimates all coefficients. However, after each M-step, the number of unknown parameters to be estimated in the next iteration is reduced according \eqref{mstep5} by using $\gb^{(t+1)}_{\mathrm{tr}}$ and $\mathbf{T}_{\mathrm{tr}}$ in the update formula of the E-step \eqref{estep_end}.
%#############################################################################
%                               section Decoding Method
%#############################################################################
\section{decoding method and implementation issues}
\label{decoding}

According to equation \eqref{brace}, we make use of the information on transmitted symbols,
obtained from the decoder, to update the channel estimate at each iteration. Besides, the decoder
requires an estimate of the channel in order to provide the probability of encoded bits. Hence,
the semi-blind channel estimation algorithm is naturally combined with the process of data decoding.
The {\it a posteriori} probability of the unknown symbol $S_k$, $p(S_k|Y_k,\widehat{H}_k^{(t)})$,
is calculated using the {\it a posteriori} probabilities provided by the decoder at the end of the $t$-th iteration as
\begin{equation}
\label{decod1}
p(S_k|Y_k,\widehat{H})= \prod_{i=1}^B P_{\rm dec}(c_{k,i})
\end{equation}
where $P_{\rm dec}(c_{k,i})$ is the {\it a posteriori} probability
corresponding to the $i$-th bit of $S_k$, $c_{k,i}$. At the first iteration,
where no {\it a priori} information is available on bits $c_{k,i}$,
$P_{\rm dec}(c_{k,i})$ are set to $1/2$.
\begin{figure}[!htb]
\centering
\psfrag{A}{$\mathbf{\Theta}$}
\psfrag{B}{\hspace{-.75em}\parbox{9.5em}{\centering \small{EM-MAP Estimation\\ of Channel Wavelet\\ Coefficients}}}
\psfrag{I}{$\gb^{(t)}$}
\psfrag{H}{\!\!$\gb_{\rm tr}^{(t)}$}
\psfrag{C}{\hspace{-.75em}\parbox{7.5em}{\centering {\small{Soft Demapping\\ \&\\ SISO Decoder}}}}
\psfrag{K}{\scriptsize  $P_{\rm dec}^{(t)}(c_{k,i})$}
\psfrag{L}{\scriptsize Uncoded Bits Probabilities}
\psfrag{M}{\scriptsize \quad\quad at last iteration}
%\psfrag{D}{\parbox{4em}{\centering\scriptsize Observed\\ Frame}}
\psfrag{D}{\scriptsize \!\! Observed Frame}
\psfrag{E}{\scriptsize $\gb^{(0)}$}
\psfrag{F}{\scriptsize $P_{\rm dec}^{(0)}(c_{k,i})=.5$}
\psfrag{G}{\hspace{-2em} \parbox{4em}{\centering\scriptsize Observed\\ Frame}}
\psfrag{J}{\hspace{-2em} \parbox{4em}{\centering\scriptsize Decoded\\ Bits}}
\includegraphics[width=0.46\textwidth,height=4.2cm]{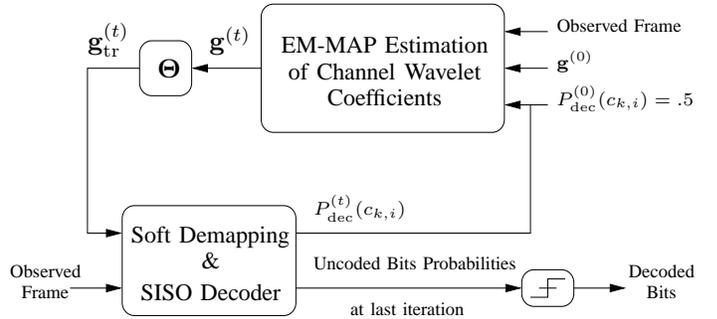}
\caption{EM-MAP channel estimation combined with the decoding process.}\label{fig2}
\end{figure}

Among several possible ways to practically implement a joint
channel estimation and decoding receiver, we adopt the following global procedure (see figure \ref{fig2}).
\begin{itemize}
\item Initialization ($t=0$)
                \begin{itemize}
                 \item Set all probabilities of coded bits $P_{\rm dec}(c_{k,i})$ to $1/2$ and derive $p(\mathbf{S}|\mathbf{Y},\gb^{(0)})$ according to \eqref{decod1}.
                 \item Initialize the unknown vector $\gb$ by $\gb^{(0)}$ obtained from pilot symbols.
               \end{itemize}

\item {\bf for} $t=1,\ldots,t_{\rm max}$
                \begin{itemize}
                \item Use the current estimate $\gb^{(t)}$ to calculate $\gb^{(t+1)}$ according to \eqref{mapstep5}.

                \item Discard the wavelet coefficients that are replaced by zero for the next iteration by evaluating $\gb^{(t)}_{\mathrm{tr}}$ and ${\mathbf{T}}_{\mathrm{tr}}$ from \eqref{mstep5}.

                \item {\bf if} $t\neq t_{\rm max}$: Use the current estimate $\gb^{(t)}_{\mathrm{tr}}$ to update the probability of encoded bits $P_{\rm dec}(c_{k,i})$ and derive $p(\mathbf{S}|\mathbf{Y},{\gb}^{(t)})$ from \eqref{decod1}.
                 \\
                {\bf else}: Decode the information data by thresholding the uncoded bit probabilities with 1/2.
                \end{itemize}
\end{itemize}
%########################################################################
%                               section simulation
%########################################################################
\section{Simulation results}
\label{simul}
In this section we present a comparative performance study of the proposed EM-MAP algorithm.
The binary information data are encoded by a non-recursive non-systematic convolutional encoder
with rate $R=1/2$ and constraint length 3. %Monte Carlo simulations are run and averaged over the transmission of at least 10000 frame.
Each frame has a payload of 1 KB along with 3 pilot symbols at the
beginning for initializing the channel of each subband. The
interleaver is random and operates over the entire frame. Among
different wavelet families, ``symmetric'' wavelet basis functions
\cite{mallat} providing the sparser representation \cite{sad1} have
been considered.Unless otherwise mentioned, the curves are obtained after $t_{\rm{max}}=4$ iterations.

First, a sparse channel model where only 20 wavelet coefficients out of total 96 have non
zero values, is considered. The second channel, referred to as Corridor, is a line of
sight (LOS) scenario issued from realistic UWB indoor channel
measurements \cite{serge} where the receive and transmit antennas
are located in a corridor separated by 9 meters.

Performance comparison is made with two pilot-only based
approach using ML and minimum mean square error (MMSE) channel
estimation, referred to as pilot-ML and pilot-MMSE. We also compare
the proposed algorithm with two semi-blind channel estimation based
on the EM algorithm, called respectively EM-Freq and EM-Wav. The first approach, consists of
estimating the channel over all of the three
subbands, using the model \eqref{eq:obs2}, similar to
\cite{Kobayashi2004a} while the second scheme is a wavelet domain EM based estimation of the channel where
the prior model is set to have a uniform distribution.

\begin{figure}[!htb]
\centering
\includegraphics[width=0.48\textwidth,height=0.3\textheight]{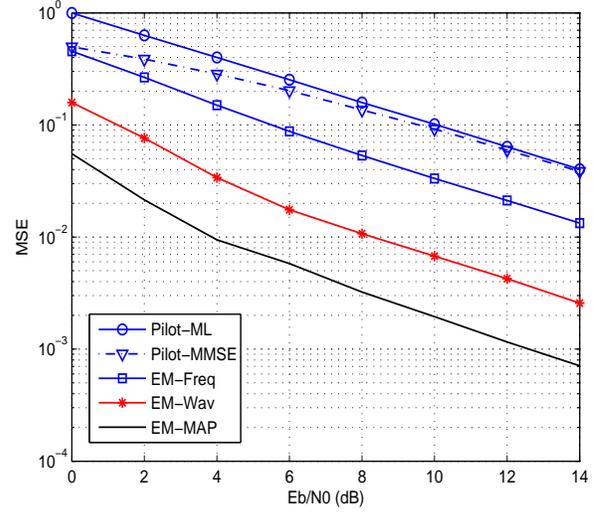}
\caption{Mean square error between the true and estimated
coefficients for the sparse channel model.}\label{mse_sparse}
\end{figure}

Figure \ref{mse_sparse} depicts the mean square error (MSE) between true and estimated channel as a function of $E_b/N_0$. It can
be noticed that, although the pilot-MMSE approach improves the
estimation accuracy for low SNR values, the performance of pilot
based channel estimation methods are very far from the family of
semi-blind methods. Comparing the wavelet domain semi-blind approach
(EM-Wav) and the frequency domain approach (EM-freq), shows that
significant gain is achieved by the former method. As
shown, the best performance is achieved by the EM-MAP method. We see
that by using EM-MAP, a  gain of almost 4 dB in SNR is achieved at
MSE=$2\times10^{-3}$, as compared to the EM-Wav method. This clearly
shows the adequacy of the EM-MAP method for the case where the
unknown channel has few non zero wavelet coefficients,
which is in perfect agreement with the prior model.

\begin{figure}[!htb]
\centering
\includegraphics[width=0.48\textwidth,height=0.3\textheight]{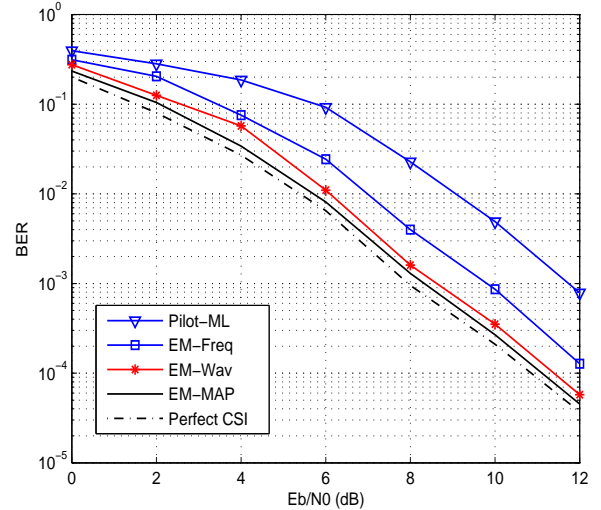}
\caption{BER performance of the EM-MAP method over the sparse channel model.}\label{ber_sparse}
\end{figure}

Figure \ref{ber_sparse} shows the BER results along with the BER for
the case of perfect channel state information (CSI). It can be seen
that at a BER of $10^{-3}$, the pilot-ML and the EM-Freq approaches
are respectively $3.9$ and $2$ dB of SNR far from the BER obtained
with the perfect channel. Furthermore, the performance of the
Pilot-MMSE approach is not shown since it was very close to that of
Pilot-ML. Also, we observe that wavelet based semi-blind methods
perform closely to the perfect CSI case. For example, at
BER=$10^{-4}$, the EM-MAP and EM-Wav method have respectively about
0.2 dB and 0.5 dB of SNR degradation from the performance obtained
with perfect CSI.

\begin{figure}[!htb]
\centering
\includegraphics[width=0.48\textwidth,height=0.3\textheight]{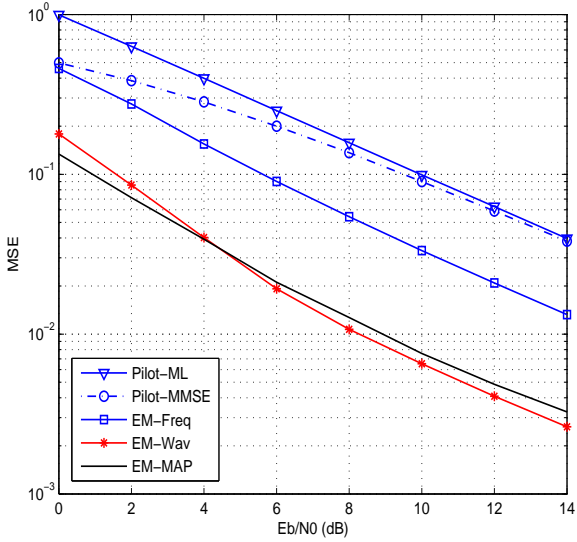}
\caption{Mean square error between the true and estimated coefficients over the Corridor channel.}\label{mse_couloir}
\end{figure}

We now evaluate the performance of EM-MAP by considering the Corridor channel. Figure
\ref{mse_couloir}, shows that
wavelet based methods again outperforms pilot based and EM-Freq
methods in terms of MSE and BER. However, the EM-MAP performance is now comparable to that of
EM-Wav method. 
This can be explained by noting that when the channel is not
sparse, small values are attributed to
$\lambda$ by the algorithm (see \eqref{parUpdate}). This leads to a gaussian prior model
with a large variance compared to the noise variance, which can be
approximated with a uniform prior. As a results, the prior becomes
less informative and the EM-MAP performs close to EM-Wav, as shown
in figures \ref{mse_couloir}. Thus, the proposed EM-MAP
algorithm is able to adapt its prior model parameters for each
propagation environment.

%\begin{figure}[!htb]
%\centering
%\includegraphics[width=0.48\textwidth,height=0.3\textheight]{./images/ber_couloir}
%\caption{BER performance of the EM-MAP method over the Corridor channel.}\label{ber_couloir}
%\end{figure}

\begin{figure}[!htb]
\centering
\includegraphics[width=0.5\textwidth,height=0.3\textheight]{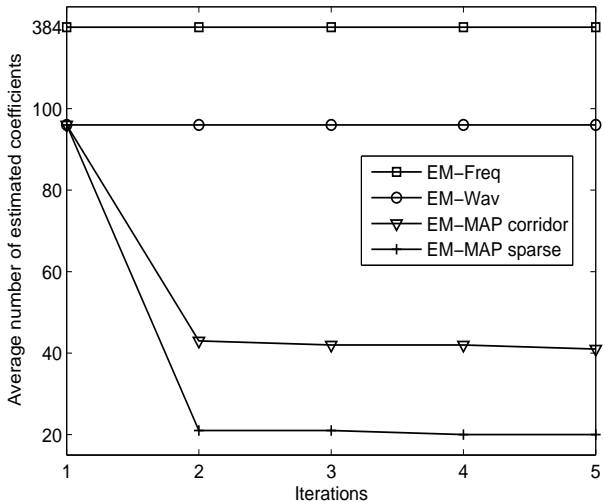}
\caption{Reduction of the number of estimated parameters through iterations, $Eb/N0=8$ dB.}\label{coef_reduc}
\end{figure}

Figure \ref{coef_reduc} shows the average number of estimated parameters versus the iteration number
different channel scenarios. As observed, the EM-MAP approach tends to reduce significantly the
number of estimated parameters. This
can be seen for the sparse channel where the number of estimated
parameters is reduced up to 20 parameters at the fifth iteration.
Furthermore, under non-sparse Corridor channel, the figure
shows that EM-MAP method is preferred to EM-Wav, due to its lower
computational load.
%#############################################################################
%                               section Conclusion
%#############################################################################
\section{Conclusion}
\label{concl}
%In this paper, we proposed a semi-blind joint channel estimation
%and data detection algorithm based on the EM algorithm that integrates the advantages of wavelet based estimation.
%By expressing the unknown UWB channel in terms of its discrete wavelet coefficients, we were able to choose a prior
%distribution that captures the possibly sparse property of UWB channels in the wavelet domain.
%This led to a MAP estimator equivalent to a hard thresholding procedure at each iteration of the EM algorithm,
%that we used to reduce the number of estimated coefficients.

%It was observed that when the channel has a sparse wavelet expansion,
%the prior model parameters which are estimated from the observed data, feed this sparsity information to the MAP algorithm.
%Moreover, we showed that in this case, the EM-MAP method provides significant reduction in the number of
%estimated parameters and outperforms all considered pilot based and semi-blind methods.
%Under non-sparse channels, although both EM-MAP and EM-Wav methods perform closely,
%EM-MAP is of advantage compared to EM-Freq and EM-Wav schemes due to its lower computational complexity.

This paper proposed a semi-blind MAP channel estimation algorithm that integrates
the advantages of wavelet based estimation.
The investigated method naturally combines the EM iterations with the decoding process.
We derived an equivalent data model for the multiband OFDM system involving
the channel over all 3 subbands expressed in the wavelet domain.
By choosing a Bernoulli-Gaussian prior distribution for the channel wavelet
coefficients, the MAP estimator yields a thresholding procedure at the M-step of the
EM algorithm which we used to reduce the number of estimated coefficients.
With only few iterations, the EM-MAP method provides significant reduction in the number of
estimated parameters and outperforms all considered pilot based and semi-blind methods.
%#############################################################################
%                               Figures at the end
%#############################################################################
%%%%%%%%%%%%%%%%%%%%%%%%%%%%%%%%%%%%%%%%%%%%%%%%%%%%%%%%%%%%%%%%%
\bibliographystyle{IEEEbib}
\bibliography{bibli_EW}
%%%%%%%%%%%%%%%%%%%%%%%%%%%%%%%%%%%%%%%%%%%%%%%%%%%%%%%%%%%%%%%%%
\end{document}

%% file: abrmath.tex
              \def\Esp#1{{\mathrm{E}}\bigcro{#1}}

\newsavebox{\fminibox}
\newlength{\fminilength}

  \def\+{^\dagger}
\def\nequiv{\not\kern-.05em\equiv}
\def\egal{\kern-.5em=\kern-.5em}        % Moins d'espace autour de "="
\def\propt{\kern-.2em\propto\kern-.2em} % Idem
\def\argmax{\mathop{\mathrm{arg\,max}}} % Mieux que \def\argmax{\arg\max}
\def\Argmax#1#2{\displaystyle \argmax_{#1}\left\{{#2}\right\}} % Ajout \HS
\def\intdouble{\int\kern-0.3em\int}
\def\inttriple{\int\kern-0.3em\int\kern-0.3em\int}

\def\rond#1{\overset{\kern-0.33em~_\circ}{#1}}
\def\rondit[#1]#2{\overset{\kern#1~_\circ}{#2}}